**Unveiling the delicate "hidden" interface conditions in $WS_2$ flakes by advanced atomic force microscopy**

Yanyan Geng[1,2,+], Chang Li[1,2,+], Shuo Mi[1,2], Manyu Wang[1,2], Xinen Han[1,2,4], Huiji Hu[1,2], Yunzhen Wang[1,2], Haojie You[1,2], Shumin Meng[1,2], Hanxiang Wu[1,2], Jianfeng Guo[3], Shiyu Zhu[3], Yanjun Li[5], Yasuhiro Sugawara[5], Sabir Hussain[6], Fei Pang[1,2], Rui Xu[1,2,*], and Zhihai Cheng[1,2,*]

[1]*Beijing Key Laboratory of Optoelectronic Functional Materials & Micro-nano Devices, Department of Physics, Renmin University of China, Beijing 100872, China*

[2]*Key Laboratory of Quantum State Construction and Manipulation (Ministry of Education), Renmin University of China, Beijing, 100872, China*

[3]*Beijing National Laboratory for Condensed Matter Physics, Institute of Physics, Chinese Academy of Sciences, Beijing 100190, China*

[4]*Experimental High School Attached to Beijing Normal University, Beijing 100032, China*

[5]*Department of Applied Physics, Gradual School of Engineering, Osaka University, Suita, Osaka 565-0871, Japan*

[6]*Tyndall National Institute, University college Cork, Lee Maltings Complex, Dyke Parade, Cork T12R5CP, Ireland*

**Abstract:** The delicate interfacial conditions and behaviors play critical roles in determining the valuable physical properties of two-dimensional materials and their heterostructures on substrates. However, directly probing these complex interface conditions remains challenging. Here, we reveal the coupled in-plane strain and out-of-plane bonding conditions in strain-engineered $WS_2$ flakes by combining dual-harmonic electrostatic force microscopy (DH-EFM) and scanning microwave impedance microscopy (sMIM). A striking contradiction is observed between the compressive-strain-induced larger bandgap (lower electrical conductivity) detected by DH-EFM, and the enhanced conductivity probed by sMIM. Comparative measurements under different sMIM modes demonstrate that this contradiction originates from a tip-loading-force-induced dynamic puckering effect, which is governed by the interfacial bonding strength. Furthermore, the progressive accumulation and subsequent release of conductivity during forward/backward sMIM-contact scans further confirms this dynamic puckering behavior, revealing pronounced differences in interface conditions between the open- and closed-ring regions of $WS_2$. This work resolves the correlation between electrical properties and interface conditions, and provides fundamental insights for interface-engineered devices.

+ These authors contributed equally to this work: Yanyan Geng, Chang Li

* Corresponding authors: ruixu@ruc.edu.cn, zhihaicheng@ruc.edu.cn

## Introduction

Two-dimensional transition metal dichalcogenides (TMDs) have emerged as promising candidates for next-generation electronic and optoelectronic devices due to their exceptional electronic [1-3], optical [4,5], and mechanical properties [6,7]. Their atomic-scale thinness enables intimate contact with substrates, but also makes their properties highly sensitive to the delicate interface conditions [8,9]. During the processes of material synthesis and integration, such as growth, transfer, and interlayer stacking [10,11], factors like thermal expansion coefficient mismatch [12-14], surface roughness [15], and bending [16,17] can introduce complex interfacial conditions. These delicate interface conditions, in turn, can effectively modify the nanoscale properties, including localized strain [18-21], charge distribution [22], band gap [23,24], and conductivity [25,26]. Consequently, these subtle interfacial effects profoundly influence critical device performance metrics, such as carrier mobility [27,28], overall electrical characteristics [29,30], and long-term stability [31]. Therefore, revealing and understanding these delicate interface conditions is not only crucial for advancing fundamental physics but also essential for engineering high-performance, reliable TMD-based devices.

To probe the interface conditions in TMD materials, a variety of experimental techniques have been employed [32-36]. Conventional techniques like Raman spectroscopy and photoluminescence measurements have been widely used to detect in-plane strain through spectral shifts [37,38], but their spatial resolution is often limited to hundreds of nanometers and cannot directly reveal the local electronic response to vertical interfacial bonding variations. Atomic force microscopy (AFM)-based methods, such as electrostatic force microscopy (EFM) Kelvin probe force microscopy (KPFM), and transverse shear microscopy (TSM), have provided valuable insights into electrical and mechanical properties [39-41]. However, these techniques are typically sensitive to either surface potential or mechanical deformation, and they struggle to distinguish between intrinsic electronic properties and extrinsic interfacial contributions. Moreover, most of them probe either in-plane or out-of-plane effects separately, without resolving their correlation. As a result, directly studying the spatial relationship between strain-induced in-plane distortions and out-of-plane interface bonding strength remains challenging.

In this study, we directly reveal the intertwined in-plane strain and out-of-plane interfacial bonding conditions in strain-engineered $WS_2$ flakes by the dual-harmonic electrostatic force microscopy (DH-EFM) and scanning microwave impedance microscopy (sMIM). DH-EFM identifies regions of intrinsic compressive strain (larger bandgap, lower conductivity), whereas sMIM paradoxically measures enhanced local conductivity in these same areas. Through comparative electrical conductivity measurements under different sMIM modes, we attribute this contradiction to the tip loading-force-induced dynamic puckering effect, modulated by the out-of-plane interfacial bonding strength. This mechanism is further confirmed by the significant enhancement and release of electrical conductivity observed during forward and backward sMIM-contact scans. Based on the above results, the in-plane strain and out-of-plane interfacial bonding strength at the open-ring and closed-ring regions of $WS_2$ flakes are schematically illustrated. Our work not only clarifies the fundamental correlation between electrical properties and interface conditions but also provides crucial guidance for the design of interface-engineered devices.

## Results and Discussion

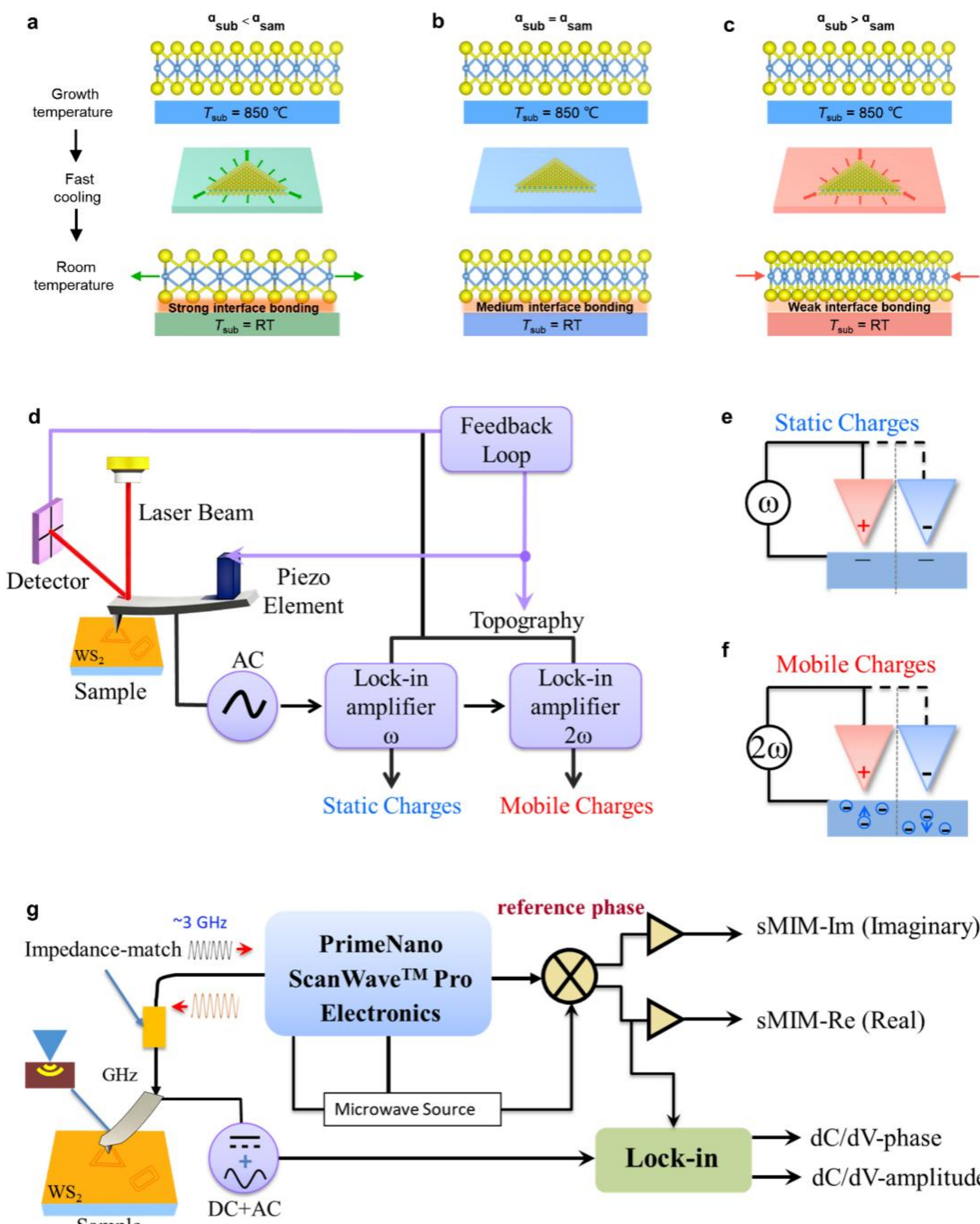

**Figure 1. Strain-engineering and electrical characterizations of $WS_2$.** (a-c) Illustration of strain-engineered interfacial conditions through thermal expansion coefficient (TEC) mismatch between substrate and sample during the CVD growth process. (d) Schematic of the DH-EFM experimental setup. The amplitude of the cantilever vibration at $f_\omega$ and $f_{2\omega}$ are obtained by the lock-in amplifiers, named as $A_\omega$ and $A_{2\omega}$. $A_\omega$ is proportional to surface potential and/or static charge, whereas $A_{2\omega}$ relates to mobile charge carriers. (e,f) The different dynamic response of static charges (e) and mobile charges (f) to the dynamic charged tip. (g) Schematic of the sMIM experimental setup. sMIM delivers a microwave signal at a few GHz to the tip apex to interact with the sample, and consequently probes its local electrical properties from analyzing the reflected microwave response.

The electrical properties of two-dimensional materials are governed by the in-plane and out-of-plane interfacial conditions between the samples and their substrates. During the chemical vapor deposition (CVD) growth, the thermal expansion coefficient (TEC) mismatch between the sample and substrate enables the realization of multiple interface conditions, including in-plane tensile, strain-free, and compressive regions, as well as out-of-plane strong, intermediate, and weak interfacial conditions, as shown in Fig. 1a-c. In this study, the $WS_2$ flakes are grown on $SiO_2$/Si substrates and rapidly cooled from the growth temperature to room temperature. Owing to the significant TEC mismatch between $WS_2$ and the substrate, nanoscale in-plane strain regions and different out-of-plane interfacial bonding strengths were generated within the $WS_2$ flakes. However, this complex interfacial condition, with in-plane strain coupling with out-of-plane interfacial bonding, poses significant challenges to conventional experimental techniques. Therefore, developing advanced experimental techniques to reveal these interfacial conditions is essential for a deep understanding of and precise control over the electrical properties of two-dimensional materials.

To probe the influence of in-plane interface conditions on the local electrical response, the DH-EFM is employed. The schematic of the DH-EFM setup is illustrated in Figs. 1d and S1. In the DH-EFM, the amplitude of the cantilever vibration at $f_{\omega}$ and $f_{2\omega}$ are obtained by the lock-in amplifiers, named as $A_{\omega}$ and $A_{2\omega}$. $A_{\omega}$ is proportional to the surface potential and/or static charges, while $A_{2\omega}$ is related to the mobile charge carriers, as shown in Fig. 1e,f. Notably, the intensity of the mobile charge carriers in $A_{2\omega}$ is inversely correlated with the local bandgap, where a larger $A_{2\omega}$ signal indicates a smaller bandgap. Therefore, analysis of the $A_{2\omega}$ images can directly reveals the electrical properties and in-plane strain distribution within $WS_2$ flakes.

To further investigate the influence of out-of-plane interface bonding on the local electrical properties, we integrated sMIM technology, as illustrated in Figs. 1g and S2. The sMIM delivers a microwave signal at a few GHz to the tip apex to interact with the sample, and consequently probes its local electrical properties from analyzing the reflected microwave response. The reflected microwave signal contains the information of the admittance/impedance of the tip-sample system, which depends on the local permittivity and conductivity of the sample. Consequently, the variation of the electronic properties results in changes of the reflected microwave signal, which is then detected by the radio frequency electronics module and processed into sMIM output, that is, real (sMIM-Re) and imaginary (sMIM-Im) components of the tip-sample impedance. By comparing sMIM signals with DH-EFM results, we can effectively decouple the contributions of in-plane strain and out-of-plane bonding, providing a comprehensive perspective on interfacial effects.

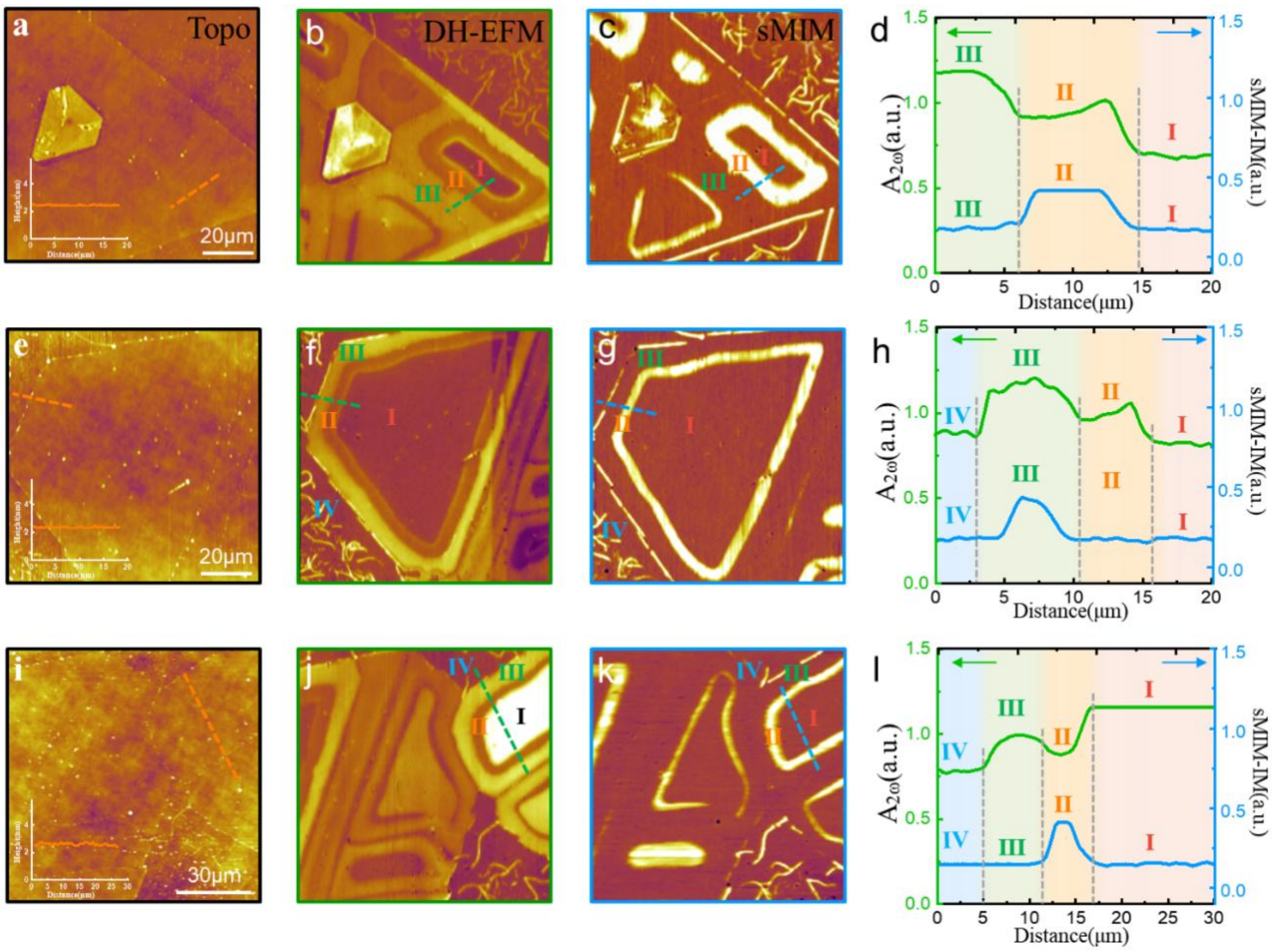

**Figure 2. Electrical characterizations of the strain-engineered $WS_2$ flakes**. (a,e,i) The AFM topography images of the $WS_2$ on the $SiO_2$/Si substrate. The insets show the line profiles along the black lines in topography images, respectively. (b,f,g) The corresponding mobile charge carrier density (MCD) images of the $WS_2$ taken by DH-EFM. The brighter/darker contrast (larger/smaller $A_{2\omega}$) correspond to a higher/lower density of mobile charge carriers and the smaller/larger band gap in the different regions. (b reproduced from [32]) (c,g,k) The corresponding electrical conductivity images of the $WS_2$ taken in the sMIM-contact measurement. The contrast in sMIM-Im reflect the conductivity of the sample with the brighter/darker contrast corresponding to higher/lower electrical conductivity. (d,h,l) The line profiles along the green and blue lines in DH-EFM and sMIM images, respectively.

The typical optical topographies of the CVD-grown $WS_2$ layers are shown in Fig. S3, where the crack boundaries along the zigzag (ZZ) orientation are clearly visible [34]. These cracks originate from in-plane tensile strain generated by the TEC mismatch between $WS_2$ and the $SiO_2$/Si substrate during cooling. Figure 2a,e,i depicts AFM topographies of $WS_2$ flakes, with no visible height difference within the flakes can be clearly resolved (insets). In our previous work [32], the corresponding mobile charge carrier density (MCD) images of the $WS_2$ are measured by DH-EFM, as shown in Fig. 2b,f,j. In the DH-EFM images, the brighter/darker contrast (larger/smaller $A_{2\omega}$) correspond to a higher/lower MCD and the smaller/larger band gap in $WS_2$. Interestingly, several nanopatterns such as the "open ring" and "closed ring" within the $WS_2$ can be clearly observed in the DH-EFM images. Analysis indicates that these open/closed ring nanopatterns (labeled as area II) correspond to the in-plane tensile strain areas (Fig. S4). The Raman experiments conducted on $WS_2$ further verify this, as presented in Fig. S5. The DH-EFM and Raman measurements can only reflect the in-plane strain distribution and cannot directly reveal the delicate "hidden" out-of-plane interfacial bonding conditions.

To further reveal the out-of-plane interface conditions, the corresponding electrical conductivity images of the $WS_2$ are obtained by the sMIM-contact mode measurement, as shown in Fig. 2c,g,k. The contrast in sMIM-Im reflect the conductivity of the sample with the brighter/darker contrast corresponding to higher/lower electrical conductivity. Notably, the sMIM-Im images display the same "open ring" and "closed ring" nanopatterns observed in the DH-EFM results. Unexpectedly, these nanopatterns (II regions) exhibit darker contrast (lower mobile charge carrier density/ lower electrical conductivity) in DH-EFM, while displaying higher electrical conductivity in sMIM at the same regions. The line profiles in Fig. 2d,h,l further confirm this inverse correlation between the DH-EFM and sMIM signals. This significant discrepancy indicates that the MCD measured by DH-EFM originates from in-plane strain, whereas the higher electrical conductivity observed in sMIM-contact mode may stems from out-of-plane interfacial bonding.

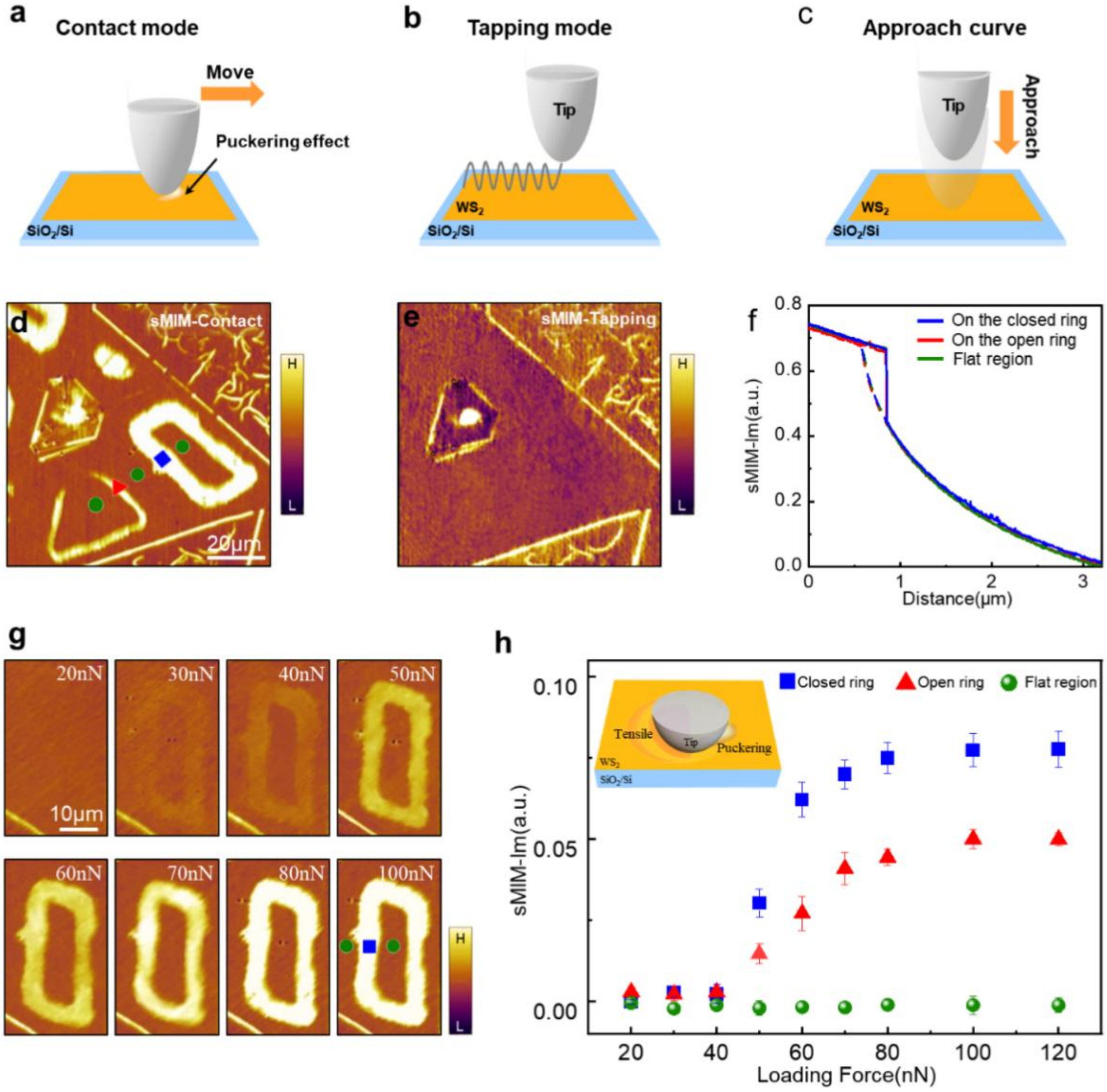

**Figure 3. Comparative electrical conductivity measurements of the strain-engineered $WS_2$ taken by sMIM measurement in different modes.** (a-c) Schematic of contact (a), tapping (b), and approaching curve (c) AFM measurement modes. (d,e) The sMIM-Im images acquired in sMIM-contact (d) and sMIM-tapping (e) measurements. The higher electrical conductivity areas are clearly resolved in (d) as the closed and open rings (marked by the blue square and red triangle, respectively), however they are not visualized in (e) taken by sMIM-tapping measurements. (f) Approaching curves of sMIM-Im measured at the locations marked by colored symbols in (d): blue square (closed ring), red triangle (open ring), and green circle (inside-ring and other flat regions), in which no difference is observed in these specific areas. (g) The acquired sMIM-Im images with different loading forces (marked in images) in sMIM-contact measurements. (h) The sMIM-Im vs. loading force curves taken at different marked locations in (d). The areas on the rings and flat regions show the increased and constant sMIM-Im signals with the loading forces in the sMIM-contact measurements. The inset shows the cartoon of puckering effect for the contact AFM scanning on the $WS_2$ film on the solid substrate.

Comparative electrical conductivity measurements of the strain-engineered $WS_2$ are investigated by sMIM measurement in different modes. The schematic of contact, tapping, and approaching curve AFM measurement modes are illustrated in Fig. 3a-c. In sMIM-contact mode (tip-sample interaction includes both vertical and lateral components), the higher electrical conductivity areas are clearly discernible as the closed and open rings, marked by the blue square and red triangle in Fig. 3d, respectively. However, in sMIM-tapping mode (tip-sample interaction only involves vertical components), these higher electrical conductivity areas are not visualized, as shown in Fig. 3e. The comparative electrical conductivity measurements in sMIM-contact and sMIM-tapping modes suggest that the electrical conductivity changes may originate from the lateral forces between the tip and sample. The approaching curves of sMIM-Im measured at the locations marked by blue square (closed ring), red triangle (open ring), and green circle (inside-ring and other flat regions) are plotted in Fig. 3f. No difference is observed in these specific areas, further confirming that the electrical conductivity increasing can only be adjusted by lateral force, rather than vertical force.

Then, sMIM-contact measurements under various loading forces are conducted to evaluate the impact of loading forces on electrical conductivity. It should be noted that at the loading force levels of ~20 nN and below, no visible higher electrical conductivity closed ring in sMIM-Im images is observed. However, the electrical conductivity increases with the loading force, and the higher electrical conductivity closed ring became clearly visible, as shown in Fig. 3g. This strength change in electrical conductivity indicates that the higher electrical conductivity of the closed ring in sMIM-contact measurements stems from the dynamic puckering effect induced tip loading force [ 42-44], rather than the intrinsic properties of the sample itself. Specifically, during sMIM-contact mode scanning, the moving tip induces puckering on the ring areas (see inset of Fig. 3h). This puckering creates a localized tensile region behind the moving tip, decreasing the band gap and enhancing the local electrical conductivity. It is noteworthy that the magnitude of the puckering effect reflects the interfacial bonding strength: weak out-of-plane interfacial bonding is more prone to generating the puckering, while the strong interlayer bonding suppresses the puckering, as illustrated in Fig. S6. The sMIM-Im vs. loading force curves taken at different marked locations are summarized in Fig. 3h. Generally, the sMIM-Im signals of the open/closed rings gradually increase with the increase of the loading forces, while the sMIM-Im signals of flat regions remain almost constant. Notably, above a loading force of ~100 nN, the sMIM-Im signal saturates, consistent with previous observations in $WS_2$ and $WSe_2/WS_2$ moiré superlattices [45,46], as shown in Fig. S7. Furthermore, the lateral force microscopy (LFM) measurements reveal the obvious differences between the closed rings and open rings areas compared to the flat regions (Fig. S8), further confirming the puckering effect. Notably, the LFM images display a distinct contrast between the inside and outside areas of the closed rings and open rings. For the open rings regions, the friction force of the inside-ring ($F_{in}$) and outside-ring ($F_{out}$) are larger than on the ring area ($F_r$): $F_{in} = F_{out} > F_r$. However, for the closed rings, the friction force order of the three areas is as follows: $F_{in} < F_r < F_{out}$. This observation is consistent with the DH-EFM signal, however, it is not visible in the sMIM signal. This discrepancy indicates that the inside and outside areas of the closed rings and open rings are in different interfacial conditions.

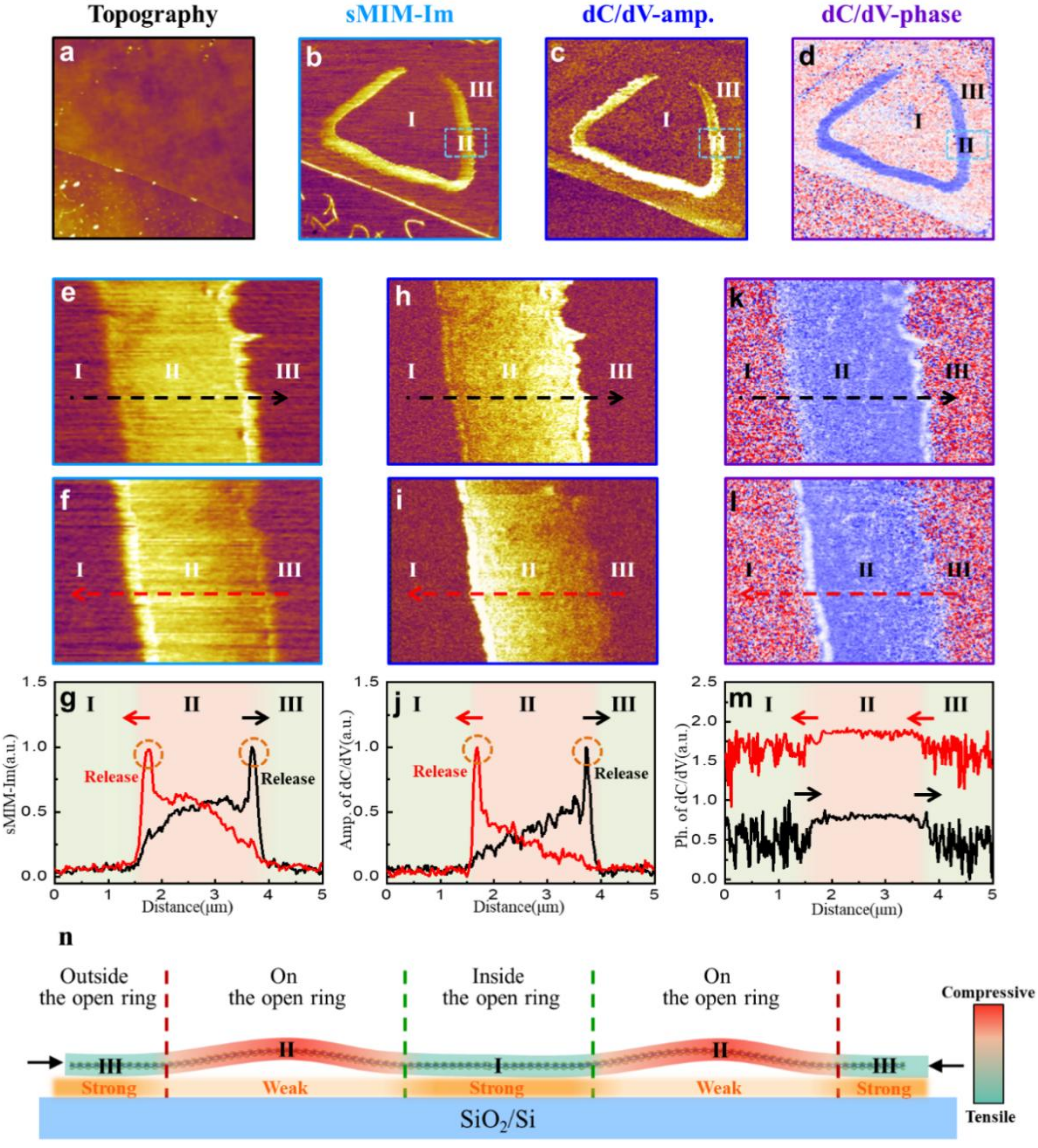

**Figure 4. The interface conditions of strain-engineered $WS_2$ at the open ring regions.** (a-d) The AFM topography (a), sMIM-Im (b), dC/dV-amplitude (c), and dC/dV-phase (d) of the open ring region. The inside/outside area of the ring and the area on the ring are marked as I/III and II, respectively. (e,h,k) The close-up sMIM-Im (e), dC/dV-amplitude (h) and dC/dV-phase (k) images taken by the sMIM-contact measurements in forward scan (marked by the black arrows). (f,i,l) The close-up sMIM-Im (f), dC/dV-amplitude (i) and dC/dV-phase (l) images taken by the sMIM-contact measurements in backward scan (marked by the red arrows). (g,j,m) The line profiles of sMIM-Im (g), dC/dV-amplitude (j), and dC/dV-phase (m) taken from the images of (e-l). (n) Schematic of the interfacial conditions for the open ring within the strain-engineered $WS_2$ flakes. The areas on the open ring are under in-plane compressive stress and out-of-plane weak interfacial bonding with the substrate. The inside and outside areas of the open ring are relatively under in-plane tensile stress and out-of-plane strong interfacial bonding with the substrate. The schematic not drawn to scale. Scan size: 40 μm × 40 μm (a-d); 5 μm × 5 μm (e-l).

To characterize variations in interfacial conditions, multimodal AFM measurements are conducted in distinct areas of the open ring, as illustrated in Fig. 4. Figures 4a-d display the AFM topography, sMIM-Im, dC/dV-amplitude, and dC/dV-phase images of the open ring region. The inside, outside, and on-ring areas are labeled I, III, and II, respectively. It can be clearly seen that the sMIM-Im value are uniform in both the inside (I) and outside (III) areas of the ring. However, the inner and outer boundaries of the ring exhibit distinct characteristics: the inner boundary is straight, while the outer boundary is squiggly. To illustrate this difference, the close-up sMIM-Im, dC/dV-amplitude and dC/dV-phase images of ring are further taken by the sMIM-contact measurements in forward/ backward scan (marked by the black/red arrows), as shown in Fig. 4e-l. During forward/backward scanning, the sMIM-Im value gradually increases along the scan direction, reaching a maximum at the ring boundary (II), exhibiting a pronounced strengthening effect. Conversely, the sMIM-Im value abruptly decrease when tip across the boundary, as shown in Fig. 4e-g. The gradual increase and sudden decrease in conductivity (sMIM-Im) reflect the accumulation and release processes of the puckering effect, respectively. This dynamic puckering effect is further confirmed by the dC/dV-amplitude and dC/dV-phase signals, which are the differential signals simultaneously acquired in Fig. 4e,f, exhibiting a superior signal-to-noise ratio. By measuring the line profiles of sMIM-Im, dC/dV-amplitude, and dC/dV-phase taken from the images of Fig. 4e-l, the accumulated puckering distance D in the open ring region is ~200 ± 50 nm.

The out-of-plane interfacial bonding strength with the substrate can be distinguished from the sMIM-Im images on the inside/outside area of the ring and on the ring area. The out-of-plane strong interfacial bonding suppresses puckering effect, while weak interfacial bonding enhances puckering effect. Based on the above results, the schematic of the in-plane and out-of-plane interfacial conditions, and dynamic puckering effect of the open ring within the strain-engineered $WS_2$ flakes are detail illustrated in Figs. 4n and S9. The areas on the open ring II are under in-plane compressive stress and out-of-plane weak interfacial bonding with the substrate, accumulating the puckering effect. In contrast, the inside I and outside III areas of the open ring are relatively under in-plane tensile stress and out-of-plane strong interfacial bonding with the substrate, releasing the puckering effect. It demonstrates that the delicate coupled in-plane strain and out-of-plane interfacial bonding collectively modulate the electrical response in $WS_2$.

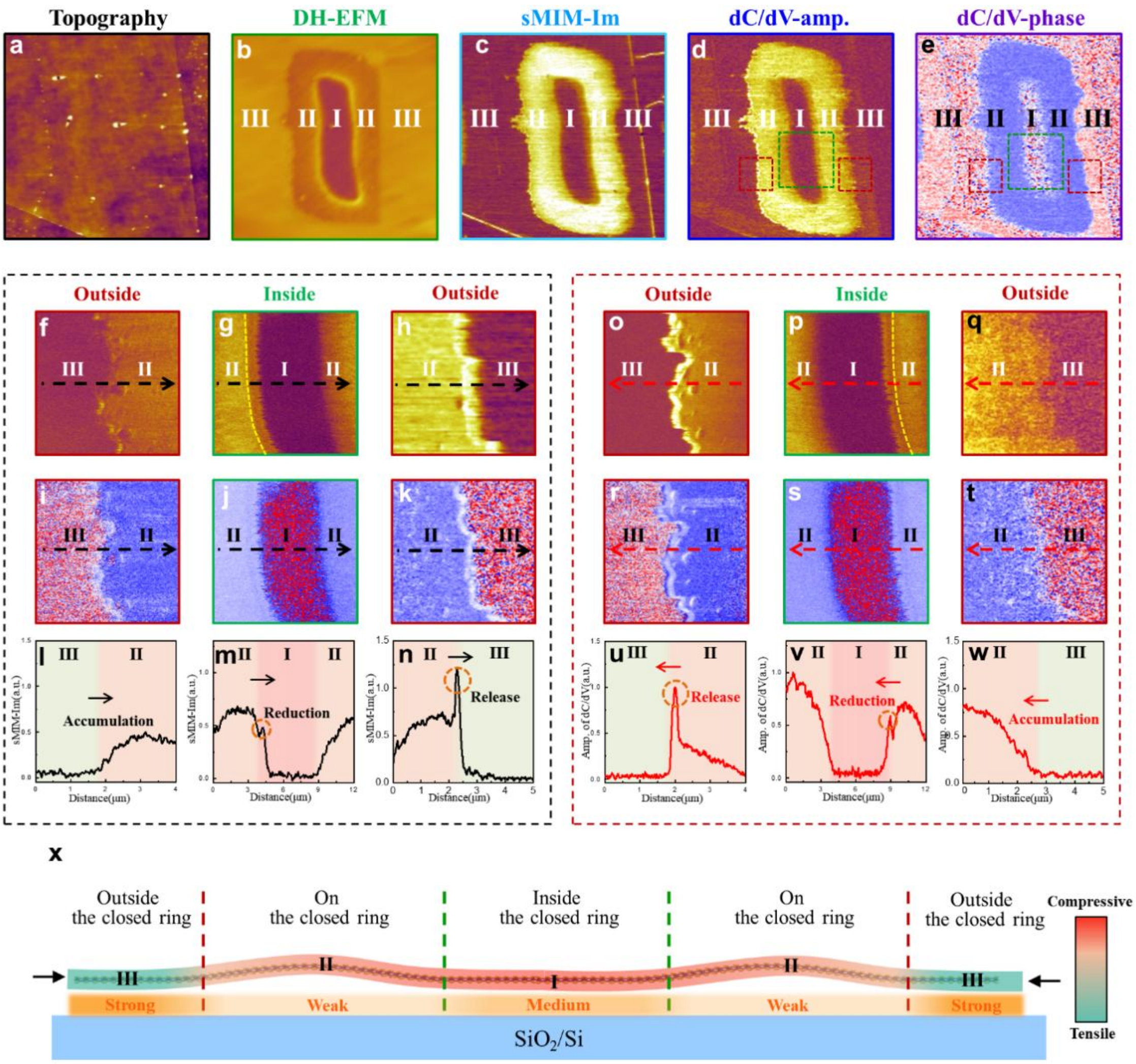

**Figure 5. The interface conditions of strain-engineered $WS_2$ at the closed ring regions.** (a-e) The AFM topography (a), DH-EFM (b), sMIM-Im (c), dC/dV-amplitude (d), and dC/dV-phase (e) of the closed ring

regions. The inside/outside area of the ring and the area on the ring are marked as I/III and II, respectively. (f-n) The close-up dC/dV-amplitude (f-h), and dC/dV-phase (i-k) images and line profiles (l-n) of the outside and inside of the closed ring taken by the sMIM-contact measurements in forward (marked by the black arrows) scans. (o-w) The close-up dC/dV-amplitude (f-h), and dC/dV-phase (i-k) images and line profiles (l-n) of the outside and inside of the closed ring taken by the sMIM-contact measurements in backward (marked by the red arrows) scans. (x) Schematic of the interfacial conditions for the closed ring within the strain-engineered $WS_2$ flakes. The areas on the closed ring are under compressive in-plane stress and weak out-of-plane interfacial bonding with the substrate. The outside area of the closed ring is relatively under tensile in-plane stress and strong out-of-plane interfacial bonding with the substrate, while the inside area is relatively under compressive in-plane stress and medium out-of-plane interfacial bonding with the substrate. The schematic not drawn to scale. Scan size: 35 μm × 35 μm (a-e); 10 μm × 10 μm (f-m); 10 μm × 10 μm (o-v).

The interface conditions at the closed ring regions are further investigated, as shown in Fig. 5. The AFM topography, DH-EFM, sMIM-Im, dC/dV-amplitude, and dC/dV-phase images of the closed ring region are displayed in Fig. 5a-e. The inside, on-ring, and outside areas are marked as I, II, and III, respectively. Pronounced contrasts in the DH-EFM signal are clearly observed between the inside and outside areas of the closed ring, whereas such contrast is not visible in the sMIM signal, suggesting distinct local strain states and interfacial coupling across the ring. To further elucidate these differences, the dynamic puckering behaviors at outside and inside of the closed ring are systematically examined using sMIM-contact measurements in forward/ backward scan (marked by the black/red arrows), as shown in Fig. 5f-n and Fig. 5o-w.

Similar to the open ring, the outer boundary of the closed ring is squiggly, while the inner boundary remains relatively straight. During the scanning, the accumulation process of the puckering effect can be clearly observed at both the outer and inner boundaries, where the sMIM-Im signal gradually increases along the scan direction and reaches its maximum at the outer boundary of the ring. However, the release process of the puckering effect, manifested as a sudden drop in the signal, is only observed at the outer boundary (Fig. 5n,u), while the inner boundary only exhibits the slight reduction in the puckering effect (Fig. 5m,v). Furthermore, the puckering accumulation distance D differs significantly between the two boundaries. The distance D of the outer boundary is same as that of open ring regions ~200 ± 50 nm, whereas D of the inner boundary of the closed ring is about 500 ± 100 nm, which is larger than that in the open ring. This implies the puckering at the inner boundary of the ring is harder than that at the outer boundary, confirming the distinct interfacial conditions between the inner and outer regions of the closed ring.

Figures 5x and S9 present a schematic illustration of the interfacial conditions in the closed ring region of the strain-engineered $WS_2$ flakes. The areas on the closed ring are under in-plane compressive stress and out-of-plane weak interfacial bonding with the substrate, favoring the emergence of the puckering effect. The outside region of the closed ring is relatively under in-plane tensile stress and out-of-plane strong interfacial bonding, suppressing the puckering effect. In contrast, the inside area is relatively under in-plane compressive stress and out-of-plane medium interfacial bonding, where the puckering effect is present but weaker than on the ring region. These results reveal that the distinct electrical responses between the inner and outer boundaries of the closed ring originate from the differences in in-plane strain and out-of-plane interfacial bonding strength, providing important insights for designing interface-tunable electronic devices.

## Conclusions

In summary, we employed the DH-EFM and sMIM to directly visualize the nanoscale interface conditions in strain-engineered $WS_2$ flakes. The apparent contradiction between the intrinsic compressive-strain-induced lower electrical conductivity in DH-EFM and the higher electrical conductivity observed by sMIM is attributed to the dynamic interfacial puckering effect governed by the interfacial bonding strength. Comparative measurements under different sMIM modes confirm that weak out-of-plane interfacial bonding facilitates the puckering-induced conductivity enhancement, while strong interfacial bonding suppresses it. By analyzing forward and backward sMIM-contact scans, we further distinguish the in-plane strain distribution and out-of-plane bonding strength in the open and closed ring regions, revealing their distinct interfacial conditions. This multimodal approach decouples the intertwined effects of in-plane strain and out-of-plane interfacial bonding on local conductivity, offering a comprehensive understanding of interfacial coupling in two-dimensional systems. Our findings provide valuable insights for interface engineering and the rational design of high-performance 2D electronic devices.

## Acknowledgments

This project was supported by the National Key R&D Program of China (MOST) (Grant No. 2023YFA1406500), the National Natural Science Foundation of China (NSFC) (No. 92477128, 92477205, 12374200, 11604063, 11974422, 12104504), the Strategic Priority Research Program (Chinese Academy of Sciences, CAS) (No. XDB30000000), the Fundamental Research Funds for the Central Universities and the Research Funds of Renmin University of China (No. 21XNLG27). Y.Y. Geng was supported by the Outstanding Innovative Talents Cultivation Funded Programs 2023 of Renmin University of China. X. E. Han was supported by the National Scientific and Technological Innovation Talent Training Program. S. H. was supported by the Science Foundation Ireland-pathway, 24/PATH-S/12757.

## Competing Interests

The authors declare no competing financial interests.

## Data Availability

The authors declare that the data supporting the findings of this study are available within the article and its Supplementary Information.

## Materials and Methods

**Growth of $WS_2$ flakes on BN.**

The $WS_2$ flakes were grown on $SiO_2$ (300 nm)/Si substrate, via traditional low pressure chemical vapor deposition (LPCVD) method. High purity $WO_3$ powders (99.5%) and sulphur powders (99.5%) applied as precursors were placed in a 25 mm quartz tube in temperature zones of 1030$^{o}$C and 180$^{o}$C, respectively. The $WS_2$ flakes were prepared for 15 min with Ar flow at a maximun pressure ~10kPa. After growth, the furnace was moved outside the sample immediately, which ensured the sample as grown rapidly cool down to room temperature.

**AFM measurements.**

The AFM (Asylum Research MFP-3D Infinity) were used under ambient condition in this paper. The introduction of used AFM technologies as follows:

**DH-EFM:** The DH-EFM measurement were performed in ambient with a home-made system, which combining the Dynamic Signal Analyzer (HF2LI, Zurich Instruments) with an Asylum MFP-3D infinity. We applied AC bias voltage with frequency ~ kHz, and then simultaneously obtain the height, $A_{\omega}$ and $A_{2\omega}$ channels. The $A_{\omega}$ channel proportional to surface potential. The $A_{2\omega}$ channel is related to mobile charge.

**sMIM**: Microwave imaging and measurements were performed in ambient with a ScanWave (Prime Nano, Inc.) sMIM add-on unit installed on the AFM. sMIM delivers a microwave signal of a few GHz to the tip apex and probes local electrical properties by analyzing the reflected microwave response.

**References:**

1. S. Manzeli, D. Ovchinnikov, D. Pasquier, O.V. Yazyev, A. Kis, Nat. Rev. Mater. **2**, 17033 (2017).

2. S. Hussain, K. Q. Xu, S. L. Ye, L. Lei, X. M. Liu, R. Xu, L. M. Xie, and Z. H. Cheng, Front. Phys. **14**(3), 33401 (2019).

3.H. Zeng, Y. Wen, L. Yin, R. Q. Cheng, H. Wang, C. S. Liu, and J. He, Front. Phys. **18**(5), 53603 (2023).

4. H.Q. Zhang, B. Abhiraman, Q. Zhang, J.S. Miao, K. Jo, S. Roccasecca, M.W. Knight, A.R. Davoyan, D. Jariwala, Nat. Commun. **11**, 3552 (2020).

5. X. Wu, H. Zhang, J. Zhang, X.W. Lou, Adv. Mater. **33**, 2008376 (2021).

6. X.M. Xu, T. Schultz, Z.Y. Qin, N. Severin, B. Haas, S.M. Shen, J.N. Kirchhof, A. Opitz, C.T. Koch, K. Bolotin, J.P. Rabe, G. Eda, N. Koch, Adv. Mater. **30**, 1803748 (2018).

7. Z.W. Peng, X.L. Chen, Y.L. Fan, D.J. Srolovitz, D.Y. Le, Light Sci. Appl. **9**, 190 (2020).

8. H. Rokni, W. Lu, Nat. Commun. **11**, 5607 (2020).

9. Z.H. Dai, N.S. Lu, K.M. Liechti, R. Huang, Curr. Opin. Solid State Mater. Sci. **24**, 100806 (2020).

10. R.Q. Ai, X.M. Cui, Y. Li, X.L. Zhuo, Nano-Micro Lett. **17**, 104 (2025).

11. J. Ji, H.-M. Kwak, J. Yu, S. Park, J.-H. Park, H. Kim, S. Kim, S. Kim, D.-S. Lee, H.S. Kum, Nano Converg. **10**, 19 (2023).

12. A. Castellanos-Gomez, R. Roldán, E. Cappelluti, M. Buscema, F. Guinea, H.S.J. van der Zant, G.A. Steele, Nano Lett. **13**, 5361-5366 (2013).

13. Y.M. Han, M.-Y. Li, G.-S. Jung, M.A. Marsalis, Z. Qin, M.J. Buehler, L.-J. Li, D.A. Muller, Nat. Mater. **17**, 129-133 (2018).

14. M.-Y. Li, Y.M. Shi, C.-C. Cheng, L.-S. Lu, Y.-C. Lin, H.-L. Tang, M.-L. Tsai, C.-W. Chu, K.-H. Wei, J.-H. He, W.-H. Chang, K. Suenaga, L.-J. Li, Science **349**, 524-528 (2015).

15. E. Blundo, T. Yildirim, G. Pettinari, A. Polimeni, Phys. Rev. Lett. **127**, 046101 (2021).

16. H.Y. Dong, S.Y. Li, S. Mi, , J.F. Guo, Z.X. Suonan, H.X. Wu, Y.Y. Geng, M.Y. Wang, H.W. Xu, L. Guan, F. Pang, W. Ji, R. Xu, Z.H. Cheng,Front. Phys. **19**, 63201 (2024).

17. J. Quereda, P. San-Jose, V. Parente, L. Vaquero-Garzon, A.J. Molina-Mendoza, N. Agraït, G. Rubio-Bollinger, F. Guinea, R. Roldán, A. Castellanos-Gomez, Nano Lett. **16**, 2931-2937 (2016).

18. W.H. Chae, J.D. Cain, E.D. Hanson, A.A. Murthy, V.P. Dravid, Appl. Phys. Lett. **111**, 143106 (2017).

19. G.H. Ahn, M. Amani, H. Rasool, D.H. Lien, J.P. Mastandrea, J.W. Ager III, M. Dubey, D.C. Chrzan, A.M. Minor, A. Javey, Nat. Commun. **8**, 608 (2017).

20. S. Hu, X. Wang, L. Meng, X. Yan, J. Mater. Sci. **52**, 7215-7223 (2017).

21. A.R. Khan, T. Lu, W. Ma, Y. Lu, Y. Liu, Adv. Electron. Mater. **6**, 1901381 (2020).

22. K. Cho, J. Pak, S. Chung, T. Lee, ACS Nano **13**, 9 (2019).

23. K.P. Dhakal, S. Roy, H. Jang, X. Chen, W.S. Yun, H. Kim, J.D. Lee, J. Kim, J. Ahn, Chem. Mater. **29**, 5124-5133 (2017).

24. S.B. Desai, G. Seol, J.S. Kang, H. Fang, C. Battaglia, R. Kapadia, J.W. Ager, J. Guo, A. Javey, Nano Lett. **14**, 4592-4597 (2014).

25. A.P. Nayak, Z. Yuan, B. Cao, J. Liu, J. Wu, S.T. Moran, T. Li, D. Akinwande, C. Jin, J. Lin, ACS Nano **9**, 9117 (2015).

26. Z.-Y. Zheng, Y.-H. Pan, T.-F. Pei, R. Xu, K.-Q. Xu, L. Lei, S. Hussain, X.-J. Liu, L.-H. Bao, H.-J. Gao, W. Ji, Z.-H. Cheng, Front. Phys.**15**(6), 63505 (2020).

27. D.D. Xu, A.F. Vong, M.I.B. Utama, D. Lebedev, R. Ananth, M.C. Hersam, E.A. Weiss, C.A. Mirkin, Adv. Mater. **36**, 2314242 (2024).

28. S.X. Yang, C. Wang, H. Sahin, H. Chen, Y. Li, S.-S. Li, A. Suslu, F.M. Peeters, Q. Liu, J.B. Li, S. Tongay, Nano Lett. **15**, 1660-1666 (2015).

29. E. Scalise, M. Houssa, G. Pourtois, V. Afanas'ev, A. Stesmans, Nano Res. **5**, 43-48 (2012).

30. P. Chaudhary, H. Lu, M. Loes, A. Lipatov, A. Sinitskii, A. Gruverman, Nano Lett. **22**, 1047-1052 (2022).

31. M. Chen, J. Xia, J. Zhou, Q. Zeng, K. Li, K. Fujisawa, W. Fu, T. Zhang, J. Zhang, Z. Wang, X. Jia, M. Terrones, Z.X. Shen, Z. Liu, L. Wei, ACS Nano **11**, 9191-9199 (2017).

32. R. Xu, Y.Z. Lun, L. Meng, F. Pang, Y.H. Pan, Z.Y. Zheng, L. Lei, S. Hussain, Y.J. Li, Y. Sugawara, J.W. Hong, W. Ji, Z.H. Cheng, J. Phys. Chem. C **126**, 16 (2022).

33. F. Pang, F.Y. Cao, L. Lei, L. Meng, S.L. Ye, S.Y. Xing, J.F. Guo, H.Y. Dong, S. Hussain, S.Z. Gu, K.Q. Xu, Y.J. Li, Y. Sugawara, W. Ji, R. Xu, Z.H. Cheng, J. Phys. Chem. C **125**, 8696-8703 (2021).

34. L. Lei, H.Y. Dong, J.F. Guo, S.Y. Xing, Y.J. Li, Y. Sugawara, F. Pang, W. Ji, R. Xu, Z.H. Cheng, Nanotechnology **32**, 465703 (2021).

35. S. Hussain, R. Xu, K.Q. Xu, L. Lei, L. Meng, Z.Y. Zheng, S.Y. Xing, J.F. Guo, H.Y. Dong, A. Liaqat, M. Iqbal, Y.J. Li, Y. Sugawara, F. Pang, W. Ji, L.M. Xie, Z.H. Cheng, Appl. Phys. Lett. **117**, 153102 (2020).

36. H.Y. Su, H.L. Zhang, J.H. Sun, H.J. Lang, K. Zou, Y.T. Peng, Nat. Commun. **15**, 9897 (2024).

37. H. Lee, Y. Koo, J. Choi, S. Kumar, H.-T. Lee, G. Ji, S.H. Choi, M. Kang, K.K. Kim, H.-R. Park, H. Choo, K.-D. Park, Sci. Adv. **8**, eabm5236 (2022).

38. S. Hussain, R. Xu, K. Q. Xu, L. Lei, S. Y. Xing, J. F. Guo, H. Y. Dong, A. Liaqat, R. Iqbal, M. A. Iqbal, S. Z. Gu, F. Y. Cao, Y. J. Li, Y. Sugawara, F. Pang, W. Ji, L. M. Xie, S. S. Chen, and Z. H. Cheng, Front. Phys. **16**(5), 53504 (2021).

39. W. Wang, Y. Zhang, Z.H. Li, L.M. Qian, Adv. Sci. **10**, 2303013 (2023).

40. Z. Y. Zheng, R. Xu, K. Q. Xu, S. L. Ye, F. Pang, L. Lei, S. Hussain, X. M. Liu, W. Ji, and Z. H. Cheng, Front. Phys. **15**(2), 23601 (2020).

41. A.N. Abramov, I.Y. Chestnov, E.S. Alimova, T. Ivanova, I.S. Mukhin, D.N. Krizhanovskii, I.A. Shelykh, I.V. Iorsh, V. Kravtsov, Nat. Commun. **14**, 5737 (2023).

42. J. Li, J. Li, J. Luo, Adv. Sci. **5**, 1800810 (2018).

43. L. Fang, D.-M. Liu, Y. Guo, Z.-M. Liao, J.-B. Luo, S.-Z. Wen, Nanotechnology **28**, 245703 (2017).

44. S. Hussain, R. Xu, K.Q. Xu, L. Lei, S.Y. Xing, J.F. Guo, H.Y. Dong, A. Liaqat, R. Iqbal, M.A. Iqbal, S. Gu, F. Cao, Y.J. Li, Y. Sugawara, F. Pang, W. Ji, L.M. Xie, S. Chen, Z.H. Cheng, Front. Phys. **16**, 53504 (2021).

45. Z.D. Chu, C.-Y. Wang, J.M. Quan, C.H. Zhang, C. Lei, A. Han, X.J. Ma, H.-L. Tang, D. Abeysinghe, M. Staab, X.X. Zhang, A.H. MacDonald, V. Tung, X.Q. Li, C.-K. Shih, K.J. Lai, Proc. Natl. Acad. Sci. U.S.A. **117**(25), 13908–13913 (2020).

46. Z.D. Chu, E.C. Regan, X.J. Ma, D.Q. Wang, Z.F. Xu, M.I. Utama, K. Yumigeta, M. Blei, K. Watanabe, T. Taniguchi, S. Tongay, F. Wang, K.J. Lai, Phys. Rev. Lett. **125**, 186803 (2020).